# Broadband-tunable spectral response of perovskite-on-paper photodetectors using halide mixing

*Alvaro J. Magdaleno, [a,b] Riccardo Frisenda,[c,d] Ferry Prins, [a,b]\* and Andres Castellanos-Gomez [c]\**

a. *Condensed Matter Physics Center (IFIMAC), Universidad Autónoma de Madrid, 28049 Madrid, Spain.*

b. *Departamento de Física de la Materia Condensada, Universidad Autónoma de Madrid, 28049 Madrid, Spain.*

c. *Instituto de Ciencia de Materiales de Madrid (ICMM-CSIC), 28049 Madrid, Spain.*

d. *Department of Physics, Sapienza University of Rome, 00185 Rome, Italy.*

AUTHOR INFORMATION

**Corresponding Author**

\* Email: ferry.prins@uam.es and andres.castellanos@csic.es

**Abstract**

Paper offers a low-cost and widely available substrate for electronics. It posses alternative characteristics to silicon, as it shows low density and high-flexibility, together with biodegradability. Solution processable materials, such as hybrid perovskites, also present light and flexible features, together with a huge tunability of the material composition with varying optical properties. In this study, we combine paper substrates with halide-mixed perovskites for the creation of low-cost and easy-to-fabricate perovskite-on-paper photodetectors with a broadband-tunable spectral response. From the bandgap tunability of halide-mixed perovskites we create photodetectors with a cut-off spectral onset that ranges from the NIR to the green, by increasing the bromide content on $MAPb(I_{1-x}Br_x)_3$ perovskite alloys. The devices show a fast and efficient response. The best performances are observed for the pure I and Br perovskite compositions, with a maximum responsivity of 376 mA/W on the $MAPbBr_3$ device. This study provides an example of the wide range of possibilities that the combination of solution processable materials with paper substrates offer for the development of low-cost, biodegradable and easy-to-fabricate devices.

**Introduction**

Paper has emerged as a promising substrate for electronic applications. The most noticeable characteristics of paper are the strongly reduced costs, biodegradation, its low mass-density, and high flexibility.[1,2] The last two provide the opportunities of creating wearable devices that can adapt to different device shapes.[3] However, fabrication of functional devices on paper remains challenging, as commonly used deposition methods for metals and semiconductors require high temperatures. In this context, the emergence of a wide range of solution processable semiconductors is of particular interest.[4,5] The use of these materials as inks has enabled high-throughput fabrication methods such as inkjet printing and roll-to-roll processing.[6,7]

Among the various classes of solution-processable semiconductors, metal-halide perovskites have emerged as a particularly versatile material for light harvesting[8–31] and light emitting applications.[32,33] Metal-halide perovskites have the general formula of $ABX_3$, in which A is a small monovalent cation (e.g. methylammonium or Cs), B is a divalent metal (e.g. Pb or Sn), and C is a halide (e.g. Cl, Br, or I). Using concentrated precursor solutions, metal-halide perovskites readily crystallize into their familiar cubic structure of corner-sharing $BX_6$ octahedra with interstitial A-site cations. Importantly, by varying the composition, broad and continuous tuning of the optical bandgap can be achieved. In the case of lead-halide perovskites, halide mixing can tune the bandgap across several hundreds of nanometers across the visible range.[34] Mixed-halide perovskites have been employed extensively to fine-tune the emission wavelength in light emitting devices,[35,36] to maximize the absorptive efficiency in tandem solar cells,[37,38] and to optimize the responsivity ($R$) in photodetectors.[9,11,19,21,31]

Indeed, previous studies have demonstrated highly efficient perovskite photodetectors with different spectral responses.[8–25] Recent studies on perovskite-on-paper photodetectors have shown the feasibility of making highly deformable and highly efficient devices. The integration of the perovskite photosensitive layers on the paper substrates was achieved by dip-coating the paper substrates in a perovskite precursor solution, drop-casting the precursor solution on the paper or by spray-coating.[39–43] More specifically, one study has demonstrated the development of a tunable photodetector by mixing different cations on the perovskite structure.[39] While some variation in the spectral response was achieved using different A-site cations,[39] these strategy does not yet make use of the much wider tunability of the optoelectronic properties of the perovskite composition using halide mixing.

Here, we integrate mixed-halide perovskites on paper substrates to develop perovskite-on-paper photodetectors that provide a broadband-tuneable spectral response. The onset of the $R$ of our devices can be shifted from the NIR to the green by increasing the %Br concentration in the $MAPb(I_{1-x}Br_x)_3$ perovskite. For the development of the devices, we combine drop-casting of the perovskite precursor solutions with the all-dry abrasion-induced deposition to print graphite electrodes on the paper.[44] Our results demonstrate the advantages of the use perovskite semiconductors as inks to achieve tailored optical responses for paper-based electronics.

**Results and discussion**

**Device fabrication**

Among the van der Waals materials, graphite is very attractive for the development of optoelectronic devices because of its lack of bandgap together with the possibility of circuit handwriting.[44] The deposition of the graphite for the device contacts can be done by rubbing pure graphite powder (MKN-CG-400, LowerFriction) against the surface of standard copy paper, as previously described by our group for van der Waals materials.[45,46] In this study, we used a vinyl stencil mask to deposit the graphite powder with the desired device layout (see **Figure S1**).

Once the device contacts were defined on the paper substrate (**Fig. 1.a.1**), $MAPb(I_{1-x}Br_x)_3$ (x = 0, 0.25, 0.5, 0.75, 1) perovskite solutions were made with 1.67 M concentration and using anhydrous n,n-dimethylformamide (DMF, Sigma, 227056-1L) as solvent (the details are listed in the ESI). The perovskite solutions were deposited by drop casting 100 μL on the area located between the graphite contacts (see **Fig 1.a.2-3**). Subsequently, the device was dried at 60ºC during 1.5 h in a $N_2$-filled glovebox (for a more detailed explanation see the ESI). The resulting devices show a perovskite coating on top of the graphite contacts (see **Fig 1.a.4**). Following this procedure, $MAPb(I_{1-x}Br_x)_3$, x = 0, 0.25, 0.5, 0.75 and 1 perovskite on paper photodetectors were created, showing clear colour variations with varying Br concentrations (see **Fig. 1.b**).

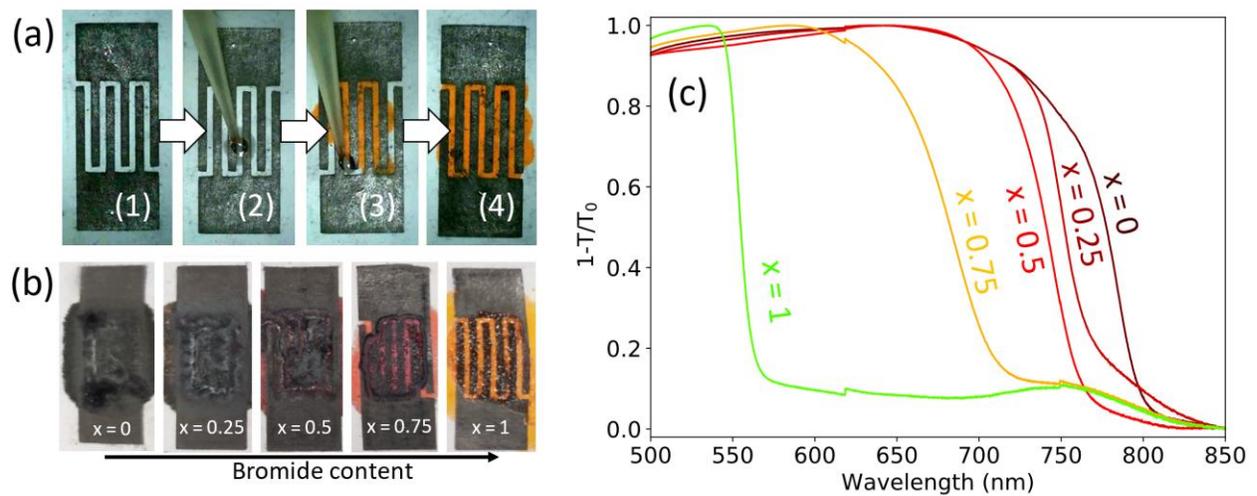

**Fig 1.** a) Device fabrication steps: Graphite electrodes on paper (1), deposition of perovskite solution (2), covering complete surface between graphite electrodes (3), drying perovskite (4). b) Resulting MAPb($I_{1-x}Br_x$)$_3$ devices (increasing %Br from left to right). c) Resulting transmission (1-transmittance) of the different perovskite-on-paper films.

**Optical characterization**

To confirm the bandgap variations of the different mixed-halide perovskites on paper, we perform optical transmissivity measurements and extract the transmission of the films. A clear tuning of the transmission spectrum is observed, as shown in **Fig. 1.c**. The transmission is plotted as 1-transmittance (1-$T/T_0$, where T is the light transmitted by the perovskite-on-paper and $T_0$ is the light transmitted by a plain paper substrate). For the pure iodide perovskite (x = 0), the transmission edge is located in the NIR. As the Br concentration (x) is increased, a gradual blue-shifting of the band edge is observed, reaching the onset of the transmission in the green for pure Bromide (x = 1).

**Morphological and compositional characterization**

To analyse the device morphology and composition along the surface, we employed scanning electron microscopy (SEM) and energy dispersive X-ray spectroscopy (EDX) using a FEI Nova NanoSEM 230. An electron energy of 18 keV was used for the EDX spectroscopy.

SEM images of the area between graphite contacts show parts that are covered by a homogeneous and mostly continuous perovskite film on the mm$^2$ scale, as can be observed in MAPbI$_3$ sample (**Fig. S2.a**). As confirmed below in the optoelectronic characterization section, this device coverage is enough to obtain a good device's functionality. Other parts of the device with a larger density of uncoated regions are also shown where the naked paper fibres are observed (see **Fig. S2.b** for MAPbI$_3$ device). The perovskite presence is confirmed by EDX analysis on the clearest regions of the SEM images (see **Fig. S2.c-e**). There, predominant Pb and I peaks are obtained for x = 0 sample (see **Fig. S2.e**). For the uncoated regions, EDX analysis shows more prominent peaks at small energies attributed to C and O. Together with the observed Ca peak, this confirms the chemical composition of standard copy paper, which is based on cellulose and calcium carbonate (see **Fig. S2.f**).

EDX analysis also confirms the presence of the corresponding perovskite compositions on the devices surface (see **Fig. S3**). For x = 0 perovskite films the EDX spectrum shows peaks at around 2.5 and 4 keV, which are related to Pb and I, respectively. On the other hand, a peak at 1.5 keV characteristic of the Br is present on the EDX spectrum of x = 1 sample instead of the I peaks, while both Br and I peaks are present on the x = 0.5 sample.

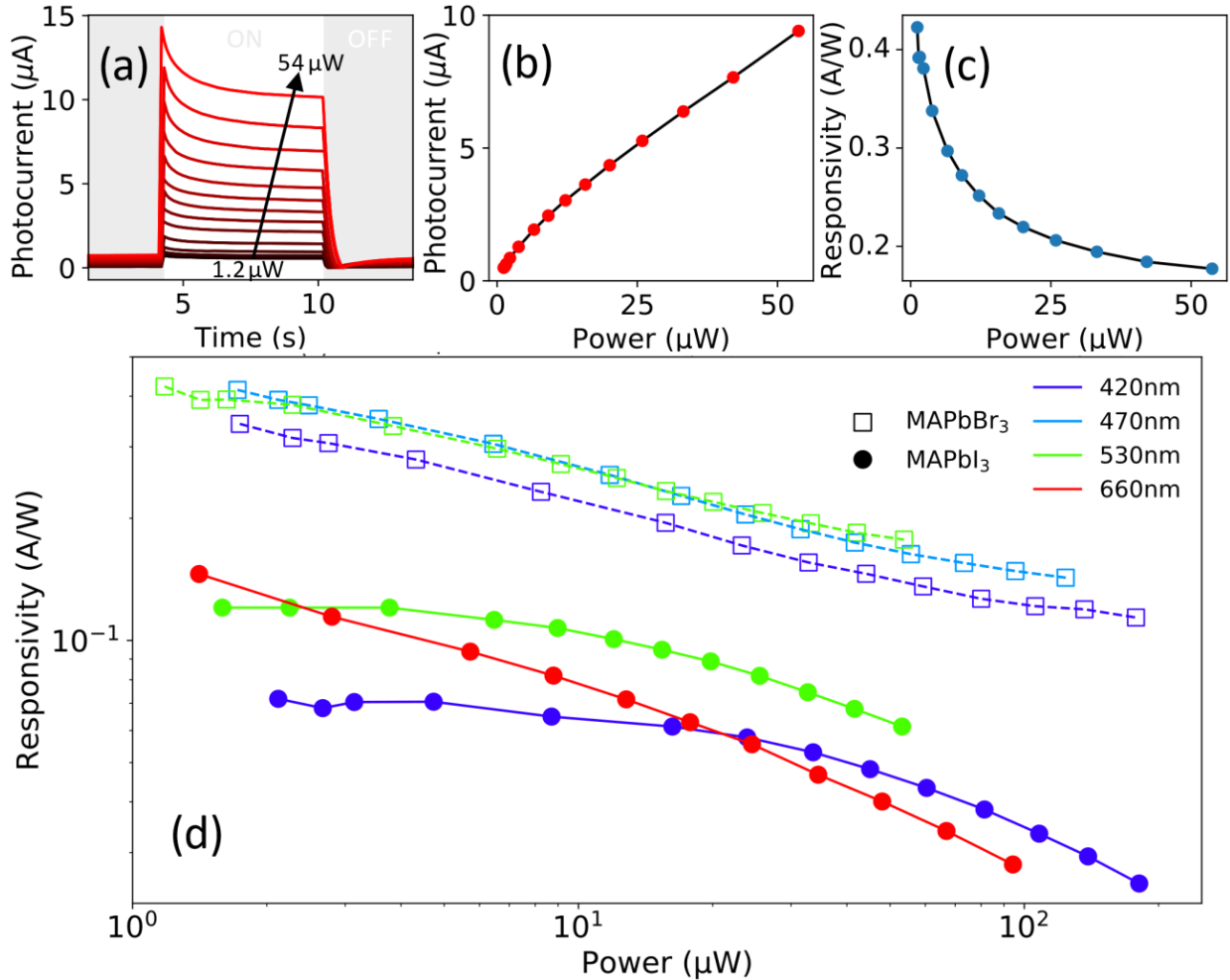

**Fig. 2.** (a) Power dependence of the photocurrents (e.g. MAPbBr$_3$ device, 530 nm LED). (b) and (c) Responsivity and photocurrent dependence on the LED power. (d) Responsivity's dependence on the LED power at different excitation wavelengths for MAPbI$_3$ (solid lines) and MAPbBr$_3$ (dashed lines).

**Optoelectronic characterization**

The performance of the perovskite-on-paper devices as photodetectors is studied by measuring their electrical transport characteristics in a home-built probe station with a source-measure unit (Keithley 2450) in the dark and upon controlled illumination. A light spot of ~71 mm$^2$ in area is used to excite the samples in all the measurements. We use high-power fiber-coupled LED sources (Thorlabs) of different wavelengths ($\lambda_{LED}$), connected to a multimode optical fiber (M28L05, Thorlabs) and a collimator (F810SMA-635, Thorlabs) to excite the devices.

Firstly, we analyse the response of our devices to different illumination intensities at fixed wavelengths and with a device bias voltage of 30 V. In **Fig. 2.a.**, as e.g. MAPbBr$_3$, excited with $\lambda_{LED}$=530 nm, we show the photocurrents generated as a function of time when switching ON and OFF the LED for increased light powers (from 1.2 to 54 µW). There is an initial sharp response when the light is turned ON (**Fig. 2.a.**, white background) with a subsequent sharp transition to an exponential decay. The transition to different regimes suggests a superposition of different physical effects. The sharp response is typically attributed to the photocurrent generation by the photoconductive effect. Other photogeneration mechanisms like photogating or bolometric could be the source of the observed slow response.[47.] To compare the device's efficiencies, we calculate the photocurrents by subtracting the background current to the current upon illumination. In addition, we calculate the R of the device following the next equation:

$$R = \frac{I_{ph}}{P} \times \frac{A_{spot}}{A_{sample}}$$

being $I_{ph}$ the photocurrent, P the illumination intensity, $A_{spot}$ the area of the incident light and $A_{sample}$ the active area of the sample that is under illumination. The photocurrent and *R* at each light power are plotted (see **Figs 2.b-c**., as e.g. MAPbBr$_3$, $\lambda_{LED}$ = 530 nm) to understand the physical origins of the responses of the device. Both variables follow a nonlinear dependence on the exciting light power. This is a common behaviour in both the MAPbI$_3$ and MAPbBr$_3$ perovskite-on-paper photodetectors. Upon illumination of different light sources, they show a non-linear trend, see **Fig. 2.d** (also **Fig. S4 and S5** to see the original photocurrents). The sub-linear power dependence suggests that minority charge carriers traps play an important role on the photocurrent generation.[47] In fact, at increasingly high power illumination there is a regime where most of the trap states are already filled-in and then trap-induced photogating effect cannot further contribute to the photocurrent generation.

Secondly, we characterize the spectral response of our MAPb(I$_{1-x}$Br$_x$)$_3$ (x = 0, 0.25, 0.5, 0.75, 1) perovskite-on-paper photodetectors. The photocurrents generated under the illumination of 14 different LED sources (Thorlabs, $\lambda_{LED}$ = 420, 455, 470, 505, 530, 565, 595, 617, 625, 660, 740, 780, 850 and 940 nm) are measured on each photodetector. The illumination intensity is kept at 0.365 mW (5 μW/mm$^2$) and the bias voltage of operation of the photodetectors is 1 V. The resulting photocurrents of each device are shown in **Fig. S6**.

When the illumination is switched ON (**Fig. S6**, white regions), all the measured photocurrents show an initial fast and sharp response whose slope slightly depends on the perovskite compositions (see **Fig. S7**). This fast response ranges from 0.1 s (x=0, 0.25 and 1) to 0.3 s (x = 0.5 and 0.75), see **Fig. S7** inset. After the fast regime, the photodetectors reach a transition towards a slower regime where the photocurrents saturate. This transition towards saturation is faster for the

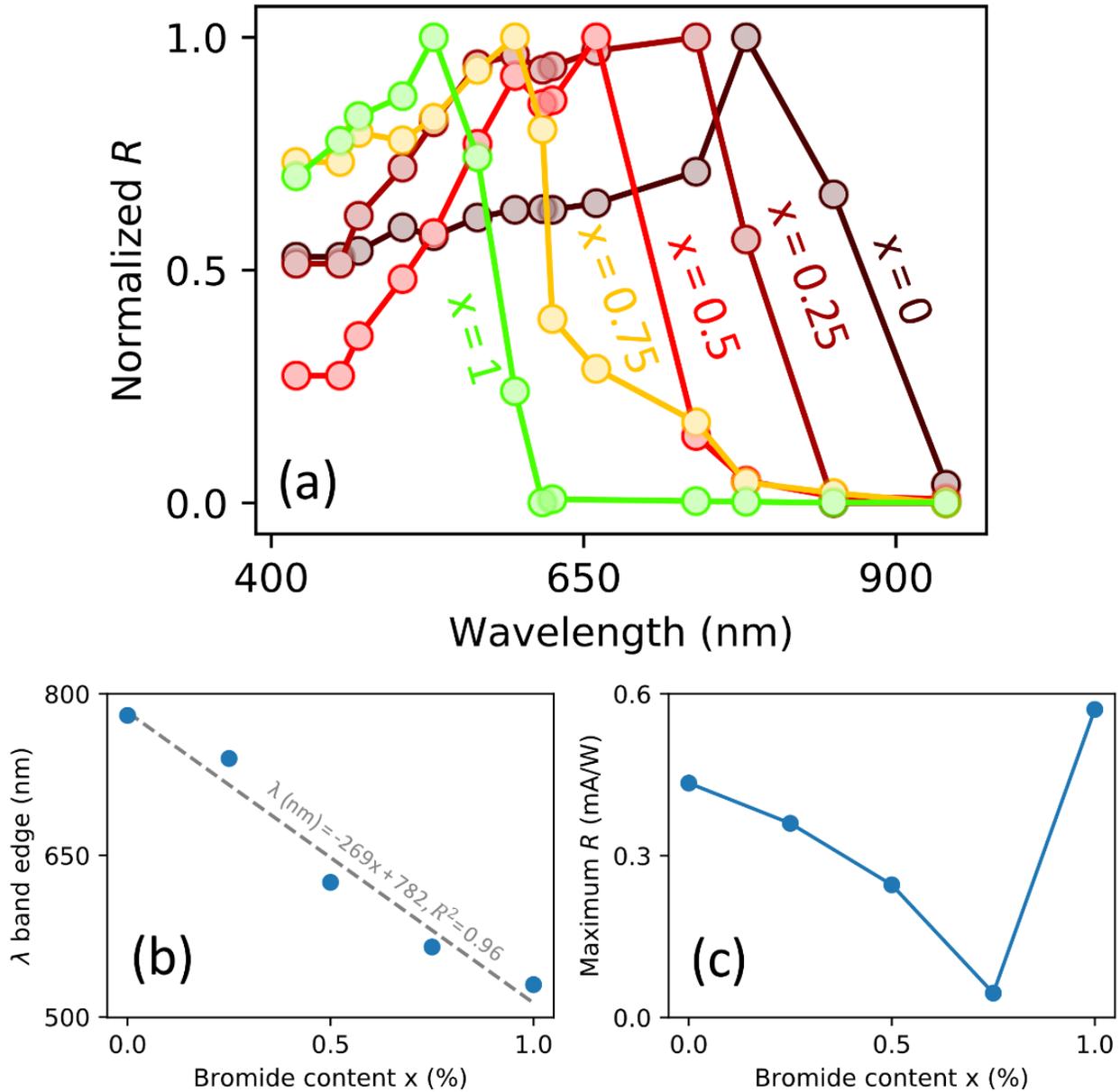

**Fig 3** a) Normalized device responsivities for each perovskite composition, b) positions of the onset of the spectral response range of the devices and c) maximum responsivity for each device compositions. All measurements attributed to this figure were done at a constant incident light power of 0.365 mW and voltage of operation of 1 V.

x = 0, 0.75 and 1 devices (less than 1 s), while for the x =0.25 and 0.5 it is more gradual (20 to 60 s, respectively). The x = 0 and 0.75 photodetectors (see **Fig. S6 and S6 inset**) show a sharp transition towards the slower regime. In these cases, the slower regime is characterized by an exponential decay of the photocurrent towards a constant value. For x = 1, the transition is less sharp but also ends with a similar exponential decay. x = 0.25 and 0.5 photodetectors show a gradual transition until they saturate at a certain photocurrent.

The photocurrent and $R$ of the devices at different wavelengths are calculated as described before in the text and plotted in **Fig. S8**. The normalized $R$ of the devices reveal a modification of the spectral range of response from the NIR to the green when increasing the Bromide content (x, %Br), see **Fig. 3.a**. All devices show a broadband response, as expected from the previously measured transmission range (see **Fig. 1.c**) and other perovskite photodetector studies.[8,18,22,48] It is important to note that the wavelength positions of the maximum $R$ have a linear dependence on the x (%Br), see **Fig. 3b**. This allows the tuning of the onset on the cut-off wavelength position of interest. We also compare the maximum $R$ of each perovskite composition (see **Fig. 3.c and S8**). They range from 0.0165 to 0.6 mA/W, for the x = 0.75 and 1 devices, respectively. The $R$ decreases in the halide-mixtures, with the minimum observed at x = 0.75. This decremental device efficiencies present on the halide-mixtures are attributed to a disordered electronic structure, as previously suggested for 3D and 2D perovskites, that can lead to electron-hole recombination.[49,50]

**Conclusions**

In conclusion, we provide a recipe for the creation of low-cost and easy-to-fabricate perovskite-on-paper photodetectors with a broadband-tunable spectral response. Taking advantage of the bandgap tunability of halide-mixed perovskites, we create perovskite-on-paper photodetectors whose spectral onset ranges from the NIR to the green, when increasing the bromide content on MAPb($I_{1-x}Br_x$)$_3$ alloys. We use the straightforward all-dry-abrasion and drop-casting approaches to print the graphite contacts and deposit the perovskite alloys, respectively. This resulted on fast device responses with efficient device performances. The best outputs are achieved on the pure I and Br devices. The study provides a new example of the wide range of properties that solution processable materials together with paper substrates offer for the development of low-cost devices.

**Author Contributions**

A.J.M., A.C.G., R.F. and F.P. led the experimental work. A.J.M. fabricated and measured the devices and carried out the data analysis with assistance from R.F., F.P., and A.C.G. R.F., A.C.G., and A.J.M. carried out the set-up preparation. F.P. and A.C.G. designed and supervised this study.

**Conflicts of interest**

There are no conflicts to declare.

**Acknowledgements**

We thank Felix Carrascoso and Wenliang Zhang (ICMM-CSIC) for support with the SEM measurements and the graphite electrodes preparation, respectively.

F.P. acknowledges support by the Spanish Ministry for Science, Innovation, and Universities through the state program (PGC2018-097236-A-I00) as well as the Ramon y Cajal program (RYC-2017-23253). A.C-G. acknowledges funding from the European Research Council (ERC) under the European Union's Horizon 2020 research and innovation program (grant agreement n° 755655, ERC-StG 2017 project 2D-TOPSENSE), the Ministry of Science and Innovation (Spain) through the project PID2020-115566RB-I00.

# Supporting information

# Broadband-tunable spectral response of perovskite-on-paper photodetectors using halide mixing

*Alvaro J. Magdaleno,[a,b] Riccardo Frisenda,[c,d] Ferry Prins,[a,b] and Andres Castellanos-Gomez[c*]*

a. Condensed Matter Physics Center (IFIMAC), Universidad Autonoma de Madrid, 28049 Madrid, Spain.
b. Department of Condensed Matter Physics, Universidad Autonoma de Madrid, 28049 Madrid, Spain.
c. Instituto de Ciencia de Materiales de Madrid (ICMM-CSIC), 28049 Madrid, Spain.
d. Department of Physics, Sapienza University of Rome, 00185 Rome, Italy

AUTHOR INFORMATION

**Corresponding Author**

* Email: ferry.prins@uam.es and andres.castellanos@csic.es

**Methods:**

**Chemicals:**

Nano graphite powder with 400 nm (PN: MKN –CG-400) average particle size was purchased at Lowerfriction Lubricants.

MAI, MABr (Greatcell Solar Materials: MS101000-10, MS301000-10), $PbI_2$, $PbBr_2$ and N,N-dimethylformamide (DMF) (Sigma Aldrich: 203602-50G, 398853-5G, 227056-1L).

**Perovskite solutions**

$MAPbX_3$ (X = I and Br) solutions were prepared by mixing MAX and $PbX_2$ with stoichiometric ratios (1:1). Subsequently, the precursor salts were dissolved in DMF by stirring at room temperature during 2 h approximately. The $MAPbI_3$ and $MAPbBr_3$ stock solutions were mixed in a (1−x)/x ratio to obtain the final solutions of $MAPb(I_{1-x}Br_x)_3$ (x = 0, 0.25, 0.5, 0.75, 1). All solutions were 1.67 M concentration and were prepared and kept in a $N_2$-filled glovebox.

**Device preparation**

a) **Printing graphite contacts**

The graphite contacts were printed applying the all-dry abrasion-induced deposition method. It is based on the rubbing of the graphite (or other van der Waals material) against the paper. The abrasion breaks the interlayer forces of the graphite and creates a film of interconnected platelets that covers the paper substrate. The rubbing was applied on graphite powder with a cotton swab on general copy paper.[1,2] See schematic of the process in **Figure S1**.

b) **Drop casting perovskite solution:**

The perovskite was deposited by drop casting 100 µL of the desired MAPb(I$_{1-x}$Br$_x$)$_3$ (x = 0, 0.25, 0.5, 0.75, 1) solution on the area between the graphite contacts (see **Fig 1.a.2-3**). The device was dried at 60ºC during 1.5 h inside of a N$_2$-filled glovebox. During the first 25 min, the paper substrate was held with to microscope slides (1 mm thick) on the borders, to avoid leakage of the perovskite solution from the contact of the paper with the hotplate. After 25 min, the paper substrate was placed in direct contact with the hotplate. After 50 min, the device was kept facing down to ensure the drying of the solution on the back side of the device. After 1 h 15 min it was flipped again to the initial position and finally left drying until 1 h 30 min (see Table S1).

| Time (min) | Action |
|---|---|
| 0 | Placing paper substrate on hotplate at 60ºC. Holding the device with a 1 mm thick glass slide on each side. Adding slowly 100 µL of perovskite solution between the contacts. |
| 25 | Leave the sample in direct contact with the hotplate |
| 50 | Flip the sample, placing the top part in contact with the hotplate |
| 75 | Flip the sample again to the initial position |
| 90 | Take the sample out of the hotplate |

**TableS1.** Steps for the development of the perovskite-on-paper devices.

**Transmission measurements.**

For these measurements we deposited the perovskite solutions on paper without graphite and let the sample dry following the same steps as described in Table S1. We used a Halogen Lamp (LHS-H100C-1) light source and the light transmitted by the resulting samples was recorded using an imaging spectrograph (Princeton Instruments, SpectraPro HRS-300, ProEM HS 1024BX3).

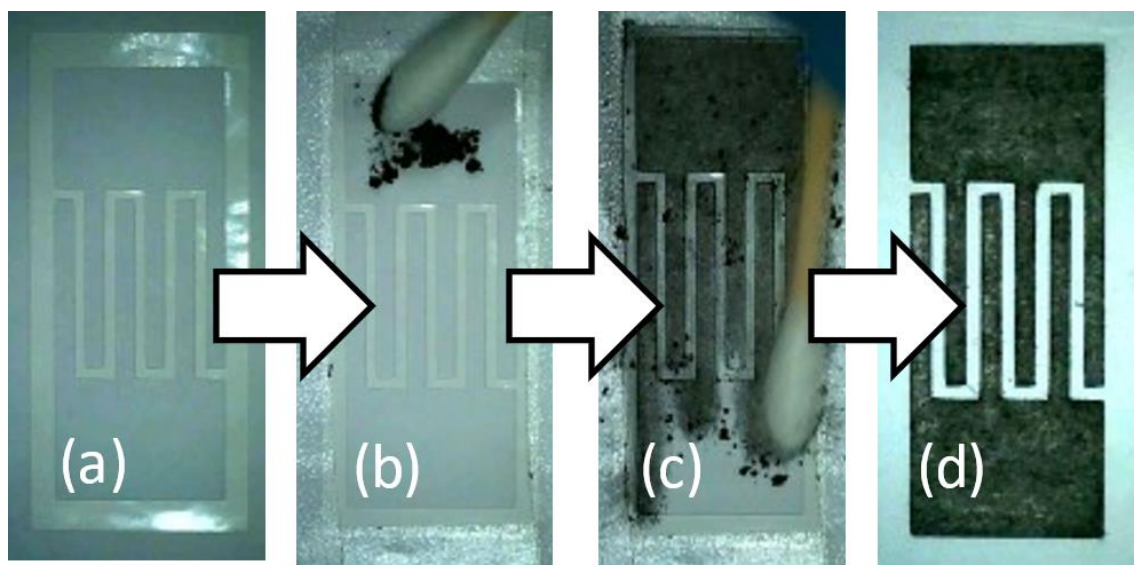

**Fig. S1** Preparation of graphite contacts on paper. (a) Copy paper with a vinyl stencil mask with device layout. (b-c) Deposition of graphite powder and rubbing until complete covering of the mask. (d) Resulting graphite on paper substrate ready for the next steps of the device preparation.

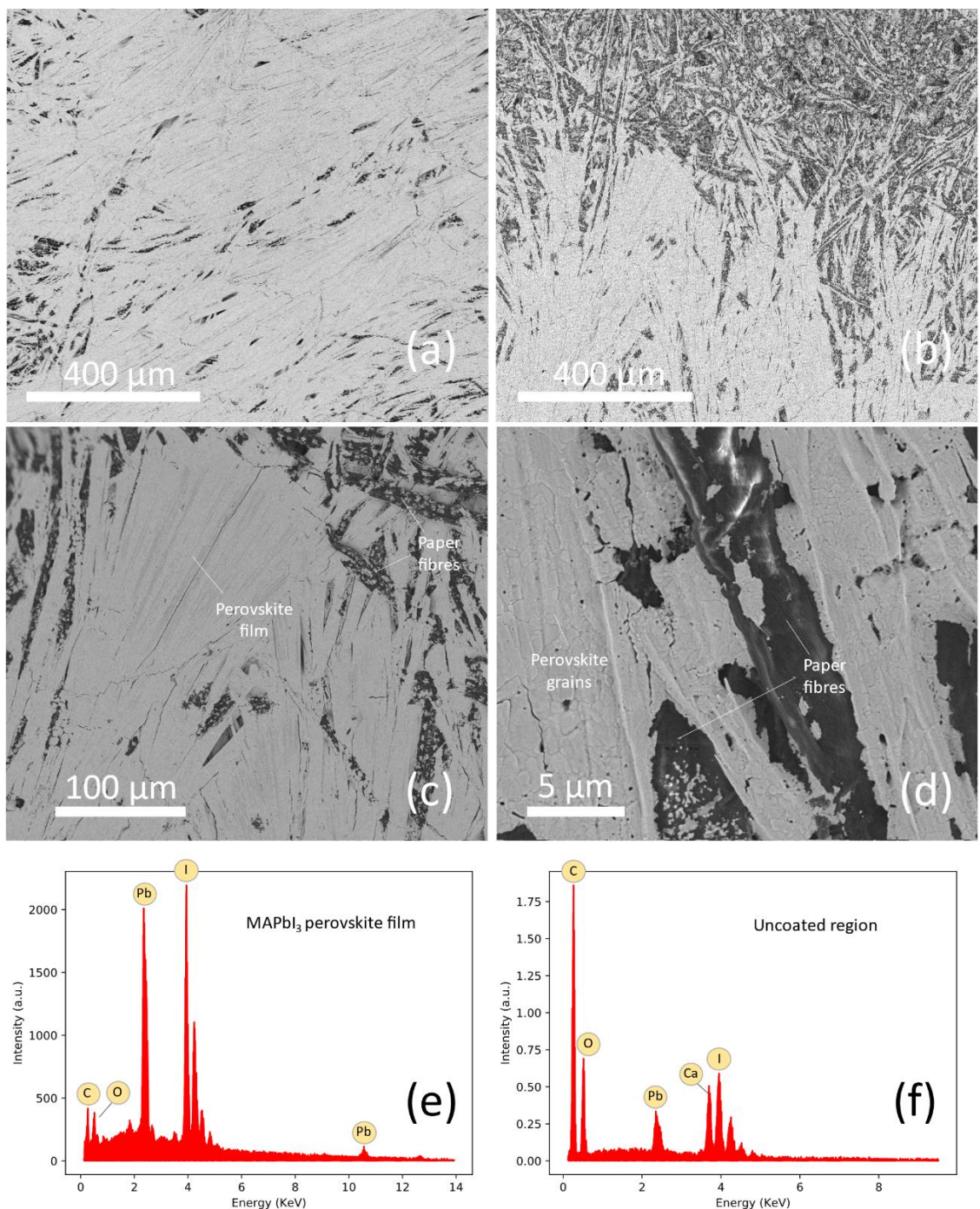

**Fig S2** SEM images and EDX results of MAPbI$_3$ perovskite-on-paper device surface. (a) Region covered by a homogeneous and mostly continuous perovskite film. (b) Place with a more predominant of uncoated paper fibres. (c-d) Zooming in the devices surface for a more detailed observation of the perovskite film and paper fibres. EDX results from the regions denoted in images (c) and (d) confirm the chemical compositions of MAPbI$_3$ (e) and of copy paper (f).

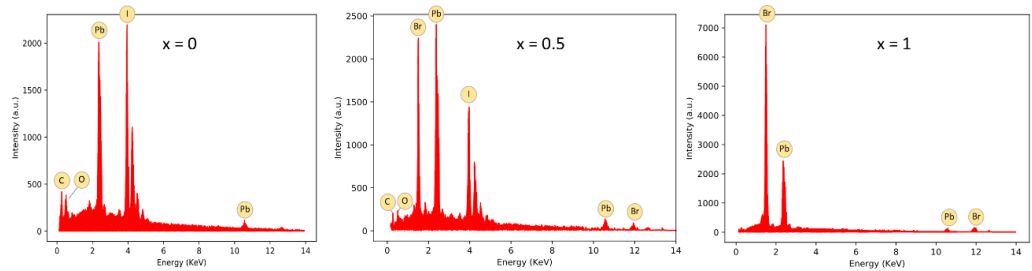

**Fig S3** EDX results of x = 0, 0.5 and 1 samples.

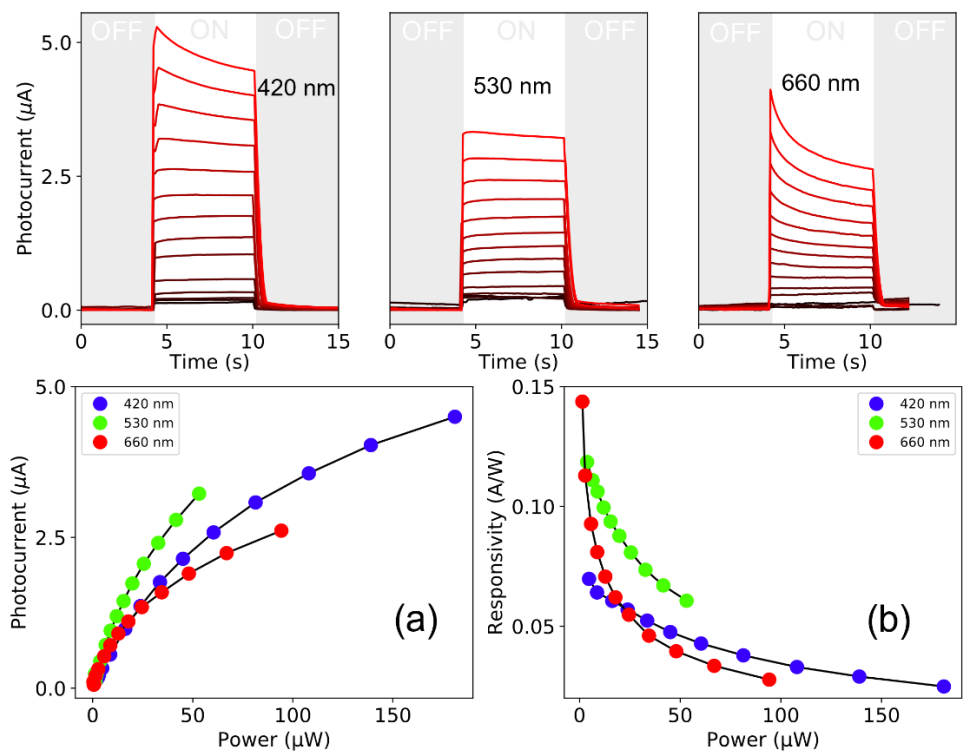

**Fig S4** MAPbI$_3$ device. Photocurrents generated at increased illumination intensities when exciting with $\lambda_{LED}$ = 420, 530 and 660 nm. (a) and (b) Photocurrent and responsivity dependence on the illumination intensities for the different $\lambda_{LED}$.

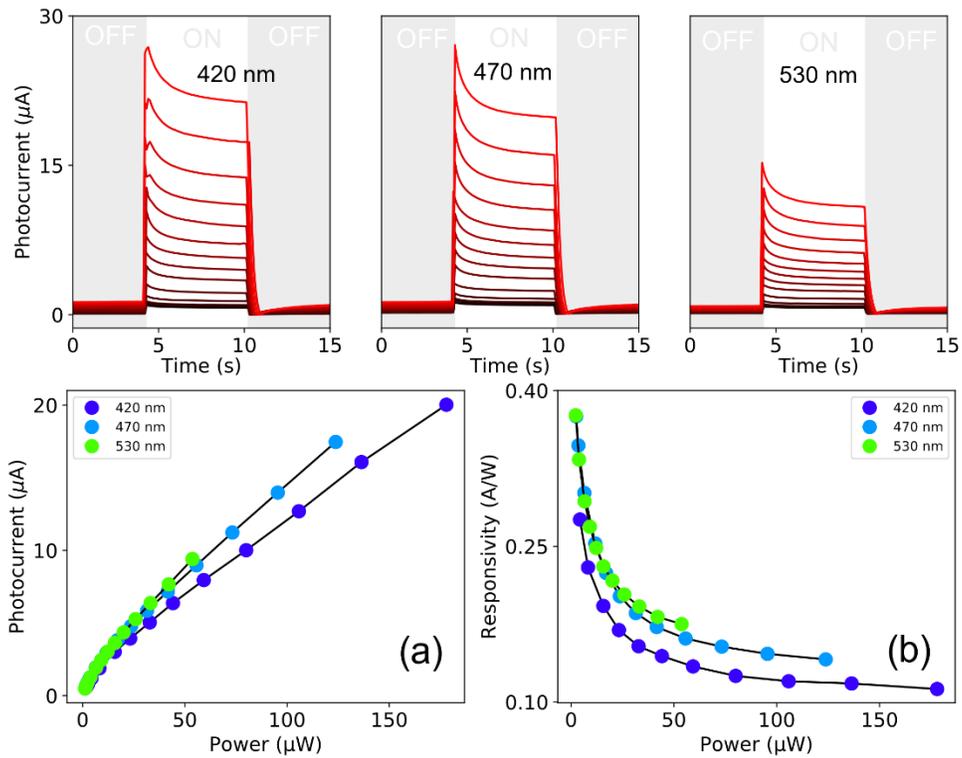

**Fig S5** MAPbBr$_3$ device. Photocurrents generated at increased illumination intensities when exciting with $\lambda_{LED}$ = 420, 470 and 530 nm. (a) and (b) Photocurrent and responsivity dependence on the illumination intensities for the different $\lambda_{LED}$.

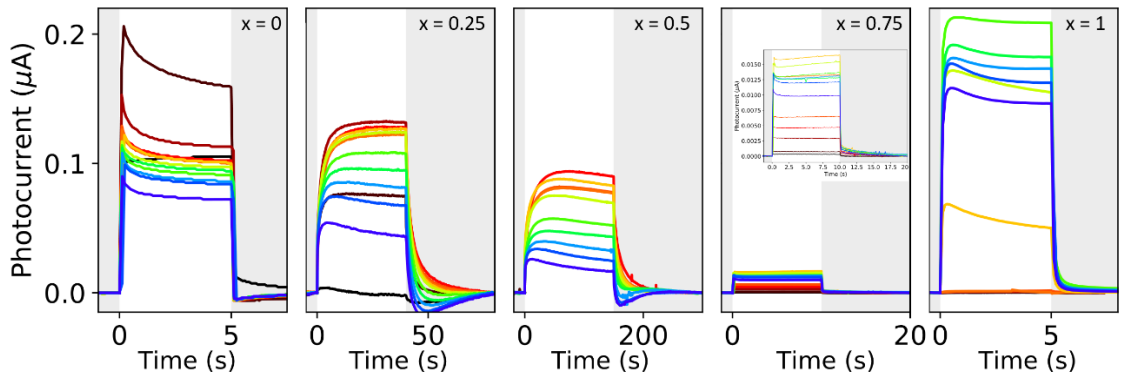

**Fig S6** Photocurrents generated on the x = 0, 0.25, 0.5, 0.75 and 1 samples when illuminating with the different LEDs. Each colour depicted is the associated to the corresponding of the incident LED light. Inset: x = 0.75 photocurrents zoomed in.

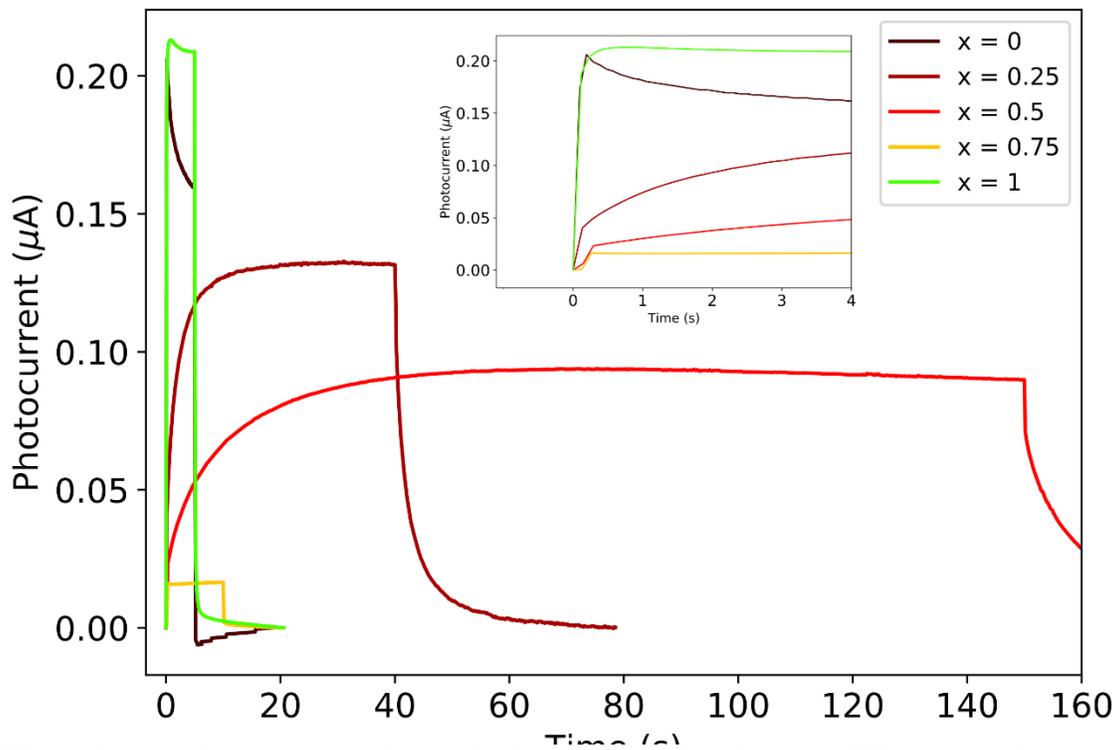

**Fig. S7** Maximum photocurrents of each device. Each device shows a different speed towards the slow regime. Each colour is the associated to the one of the incident LED light. Inset: Initial responses of the devices.

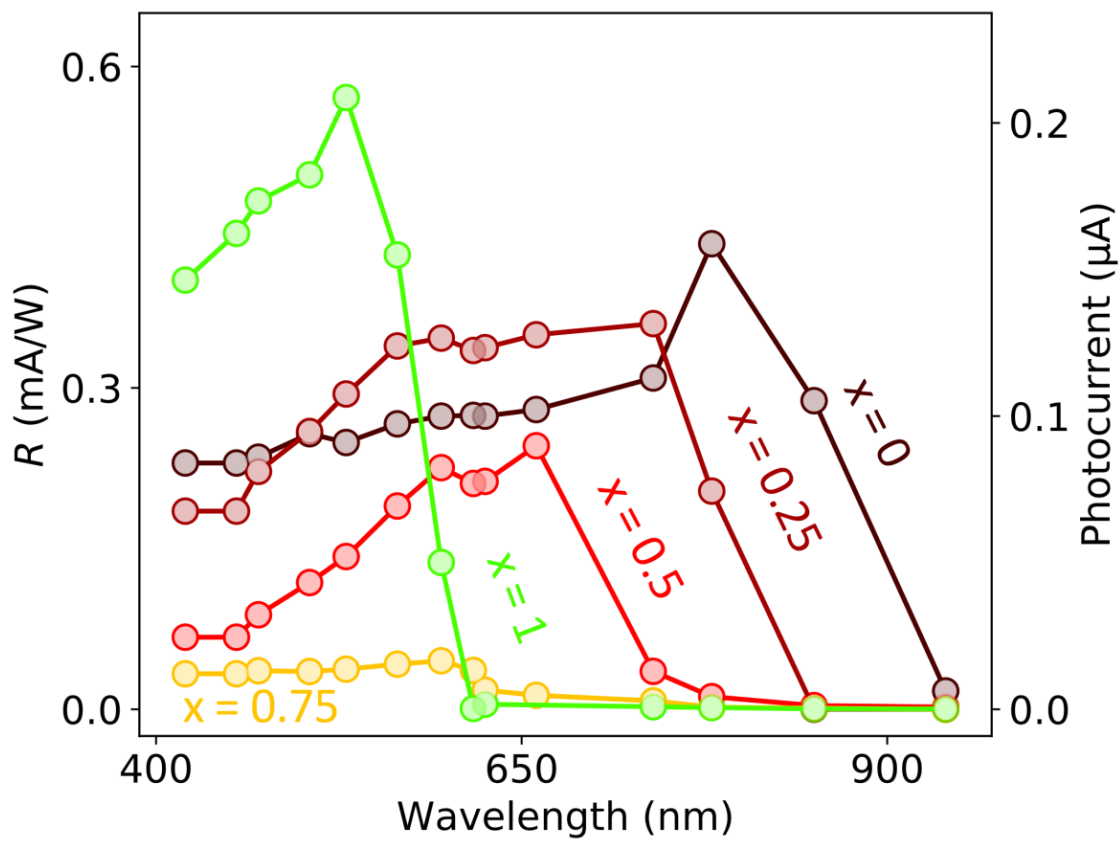

**Fig S8** Responsivities and photocurrents of the different devices when exposed to LED lights of different wavelengths at a constant illumination intensity of 0.365 mW (5 μW/mm$^2$) and operation voltage of 1 V. The depicted colours are related to the maximum *R* wavelength positions.